\documentclass[twocolumn]{jpsj3}
\usepackage{txfonts,color}
\newcommand{\vtr}[1]{\mbox{\boldmath $#1$}}

\title{A Superparamagnetic State Induced by a Spin Reorientation Transition
in Ultrathin Magnetic Films}

\author{Yousuke NORIZUKI and Munetaka SASAKI}
\inst{Department of Applied Physics, Tohoku University, Sendai 980-8579}

\abst{
We investigate a spin reorientation transition (SRT) in ultrathin magnetic films 
by Monte-Carlo simulations. We assume that the lateral size of the film is relatively small 
and it has a single-domain structure. To gain insights into the SRT, 
we measure a free-energy as a function of perpendicular and in-plane magnetizations. 
As a result, we find that the system is in a {\it superparamagnetic state} 
at the SRT temperature. The disappearance of magnetization around the SRT temperature, 
which is observed in experiments, emerges due to dynamical fluctuations in magnetization
which are inherent in the superparamagnetic state. This observation is in contrast to that 
in large ultrathin magnetic films that the disappearance of magnetization is caused by 
a static magnetic structure with many complex domains.
}

\kword{spin reorientation transition, ultrathin magnetic film, Monte-Carlo simulation}

\begin{document}
\maketitle
\section{Introduction}
\label{sec:introduction}

In the last several decades, ultrathin magnetic films have studied extensively 
due to both the fundamental interest of low-dimensional magnetism and 
the numerous perspectives of applications. These studies have revealed that 
complex magnetic order and curious phenomena are induced in ultrathin magnetic 
films by competition among several interactions such as ferromagnetic exchange 
interactions, magnetic anisotropy energies, and magnetic dipolar interactions. 
A spin reorientation transition~\cite{Papas90,Allenspach90,Allenspach92,Qiu93,Speckmann95} (SRT) 
is one of such phenomena. This is a transition from a in-plane magnetized state 
to a perpendicularly magnetized state. This transition is induced by decreasing 
the thickness of the film or by decreasing the temperature. 
The SRT has been extensively investigated both experimentally and theoretically 
until now (see reviews~\citen{DeBell00,JensenBennemann06} and references therein).

The main issue we address in the present study is the disappearance of magnetization 
around the SRT temperature $T_{\rm SRT}$~\cite{Papas90,Qiu93}. This abrupt drop in magnetization 
around $T_{\rm SRT}$ is called {\it pseudogap}~\cite{Qiu93}. 
Papas {\it et al.} proposed that there are two possible origins which cause this pseudogap. 
The first is a dynamical origin that the system is in a paramagnetic state 
because perpendicular and in-plane anisotropies compensate with each other around $T_{\rm SRT}$. 
As a result, magnetization heavily fluctuates with time, and the time average 
of each component of magnetization, which is measured in experiments, becomes zero. 
The second is a static origin that complex magnetic domains, which cause 
an almost complete loss in the total magnetization of the sample, emerge in the vicinity 
of $T_{\rm SRT}$. For large ultrathin films with the lateral size of several tens 
of micrometers, it was observed that a complex domain structure appears 
near the SRT temperature~\cite{Allenspach92,Speckmann95}. Such behavior 
was also observed in Monte-Carlo (MC) study~\cite{Carubelli08} if the corresponding 
system size is large enough~\cite{note1}. 
These results strongly indicate that the pseudogap is caused by the static origin 
for large films. However, it has not been clarified whether 
the pseudogap emerges due to the dynamical origin even if the lateral size of the film 
is not large and the system has a single-domain structure. 

To address this issue, we examine the SRT of ultrathin magnetic films 
with focusing on the pseudogap around $T_{\rm SRT}$. The corresponding size of the film 
is chosen to be relatively small so that the system has a single-domain structure. 
To gain insights into the properties of the SRT, we measure a free-energy as a function of 
the two order parameters of the SRT, i.e., perpendicular magnetization $m_{\perp}$ and 
in-plane magnetization $m_{\parallel}$. To our knowledge, this is the first time that 
such free-energy is measured in the study of the SRT. 
The free-energy is calculated numerically by using the method proposed 
in ref.~\citen{WatanabeSasaki11}. As a result, we found that the free-energy 
around $T_{\rm SRT}$ has a structure that low free-energy region widely spreads 
in the $(m_{\perp},m_{\parallel})$ space. Because the amplitude of the magnetization 
$m\equiv(m_{\perp}^2+m_{\parallel}^2)^{1/2}$ is non-zero in the low free-energy region, 
spins are aligned in the same direction. However, the direction of the magnetization heavily 
fluctuates with time as a result of the exploration inside the low free-energy region. 
In short, the system is in a {\it superparamagnetic state} at $T_{\rm SRT}$. 
Our results indicate that the pseudogap emerges due to the dynamical origin 
even in relatively small ultrathin magnetic films with a single-domain structure. 


\section{Model}
\label{sec:model}
In the present work, we consider a spin model on a $32\times 32 \times 1$ square lattice 
with open boundaries. The Hamiltonian of the model is described as
\begin{align}
 \label{Hami}
  {\cal H} &= -J\sum_{\langle i,j \rangle} \vtr{S}_i \cdot \vtr{S}_j
      +C_{\rm d}\sum_{i<j} \left( \frac{\vtr{S}_i \cdot \vtr{S}_j}{r_{ij}^3} 
        -3\frac{(\vtr{S}_i\cdot \vtr{r}_{ij})(\vtr{S}_j\cdot\vtr{r}_{ij})}{r_{ij}^5}
	\right) \nonumber \\ 
      &-C_{\rm u}\sum\nolimits_i(S_i^z)^2 ,
\end{align}
where $\vtr{S}_i$ is a classical Heisenberg spin with an absolute value of unity, 
$\vtr{r}_{ij}\equiv \vtr{r}_i-\vtr{r}_j$, $\vtr{r}_i$ is the position vector 
of a site $i$ in the unit of the lattice constant, and 
$r_{ij} \equiv | \vtr{r}_{ij} |$. 
In the right hand side of eq.~(\ref{Hami}), the first term, the second term, 
and the third term denote ferromagnetic exchange interactions, dipolar interactions, 
and magnetic anisotropy energies, respectively. The sum of exchange interactions 
runs over all of the nearest-neighboring pairs. We assume that magnetic anisotropy 
is uniaxial and its easy axis is parallel to the $z$-axis, where 
the $z$-axis is perpendicular to the film. 
This type of model has been often used in the study of 
the SRT~\cite{Carubelli08,Chui94,Hucht95,MacIsaac98,Vedmedenko99,Vedmedenko00,
Vedmedenko02,SantamariaDiep00}. 

The three parameters $J$, $C_{\rm d}$, and $C_{\rm u}$ in eq.~(\ref{Hami}) 
represent the strength of each energy term. In the present work, we performed simulations 
for four different sets of the parameters. 
The values of $C_{\rm d}/J$ and $C_{\rm u}/J$ in the four cases are shown in Table~\ref{param}. 
The ratio $C_{\rm d}/C_{\rm u}$, which is $0.14$ in all of the four cases, is chosen 
so that the SRT is clearly observed. 
In micromagnetic calculations, the three parameters in eq.~(\ref{Hami}) 
are calculated from the material parameters and the mesh size as
\begin{equation}
J=2\delta A,\quad C_{\rm d}=J_{\rm s}^2\delta^3/(4\pi\mu_0),\quad C_{\rm u}=K_{\rm u}\delta^3,
\label{eqn:parameters}
\end{equation}
where $\delta$ is the mesh size, $A$ is the exchange stiffness constant in continuous limit, 
$J_{\rm s}$ is the saturation polarization, $K_{\rm u}$ is the magnetic anisotropy constant, 
and $\mu_0$ is the vacuum permeability. The mesh size $\delta$ shown 
in Table~\ref{param} is calculated from the ratio $C_{\rm d}/J$ by assuming that 
$A = 30.2\times10^{-12}~{\rm J/m}$ and $J_{\rm s} = 1.79~{\rm T}$. 
These values are the same as the material parameters of cobalt~\cite{MicromagneticBook}, 
which is one of the representative magnetic materials. 
The sizes of the film shown in Table~\ref{param} are calculated from the mesh size 
and the lattice size. The magnetic anisotropy constant $K_{\rm u}$ is calculated from the ratio 
$C_{\rm d}/C_{\rm u}=0.14$ as $1.45\times10^6~{\rm J/m^3}$. Since the magnetic anisotropy 
constant of cobalt is $4.53\times10^5~{\rm J/m^3}$~\cite{MicromagneticBook}, 
this value is about three times larger than 
that of cobalt. In the present study, we examine the mesh-size dependence 
by comparing the data of these four cases.

\begin{table}
\caption{The four sets of parameters used in the present study. The mesh size 
$\delta$ is calculated from eq.~(\ref{eqn:parameters}) and the ratio $C_{\rm d}/J$ 
by assuming that $A = 30.2\times10^{-12}~{\rm J/m}$ and $J_{\rm s} = 1.79~{\rm T}$. 
The size is calculated from the mesh size $\delta$ and the lattice size $32\times 32\times 1$. 
}
\label{param}
\begin{center}
\begin{tabular}{ccccc}
\hline
 &\multicolumn{1}{c}{$C_{\rm d}/J$} & \multicolumn{1}{c}{$C_{\rm u}/J$} &
\multicolumn{1}{c}{$\delta~[{\rm nm}]$} & \multicolumn{1}{c}{size$~[{\rm nm}^3]$}\\ 
\hline
\ case 1 & 0.007 & 0.05 &  1.44 & $46.1\times 46.1\times 1.44$\\
\ case 2 & 0.028 & 0.2  &  2.89 & $92.5\times 92.5\times 2.89$\\
\ case 3 & 0.056 & 0.4  &  4.08 & $131\times 131\times 4.08$\\
\ case 4 & 0.070 & 0.5  &  4.56 & $146\times 146\times 4.56$\\
\hline
\end{tabular}
\end{center}
\end{table}

\section{Method}
\label{sec:method}
We next explain how we measure free-energy as a function of $m_{\perp}$ and $m_{\parallel}$. 
The free-energy is defined by 
\begin{align}
 \label{def_f}
 \exp \left[-\beta F \left( \beta; m_{\perp},m_{\parallel} \right) \right] 
   \equiv {\rm Tr}_{\{\bold{S}_i\}}
  \exp \left[-\beta {\cal H}\{\bold{S}_i\} \right] \nonumber \\
   \times \delta \left( m_{\perp}-m_{\perp}^*\{\bold{S}_i\} \right)
   \delta \left( m_{\parallel}-m_{\parallel}^*\{\bold{S}_i\} \right) ,
\end{align}
where $\beta$ is the inverse temperature and ${\cal H}\{\bold{S}_i\}$ is the 
Hamiltonian defined by eq.~(\ref{Hami}). 
The trace in the right-hand side of eq.~(\ref{def_f}) runs over all of possible configurations.
Note that the right-hand side of the equation is proportional to the 
probability that spin configurations with magnetizations 
$\left(m_{\perp},m_{\parallel}\right)$ are observed. 
$m_{\perp}^*\{\bold{S}_i\}$ and $m_{\parallel}^*\{\bold{S}_i\}$ are the perpendicular and in-plane 
magnetizations calculated from a configuration $\{\bold{S}_i\}$, respectively, i.e., 
 \begin{equation}
 \label{def_mperp}
  m_\perp^*\{\bold{S}_i\} = \frac{1}{N} \left| \sum\nolimits_i S_i^z \right| ,
 \end{equation}
 \begin{equation}
   \label{def_mpara}
   m_\parallel^*\{\bold{S}_i\} = \frac{1}{N} 
   \sqrt{ \left( \sum\nolimits_i S_i^x \right)^2 
        + \left( \sum\nolimits_i S_i^y \right)^2    },
 \end{equation}
where $N$ is the number of spins.

We used the method proposed in ref.~\citen{WatanabeSasaki11} 
to calculate $F\left(\beta;m_{\perp},m_{\parallel}\right)$ numerically. 
In this method, a variant of the Wang-Landau method~\cite{WangLandau01A,WangLandau01B} 
is combined with the stochastic cutoff (SCO) method~\cite{SasakiMatsubara08}, which was recently 
invented for long-range interacting systems. This method enables us to calculate 
$F\left(\beta;m_{\perp},m_{\parallel}\right)$ with reasonable computational time.
Furthermore, long-range dipolar interactions are taken into account without any 
approximations by the use of the SCO method. 
The detailed conditions of free-energy measurements are as follows. 
We set the initial value of the modification constant $\Delta F$ in 
ref.~\citen{WatanabeSasaki11}, which is related with the modification factor $f$ 
in the Wang-Landau method~\cite{WangLandau01A,WangLandau01B} by $f=\exp(\Delta F)$, 
to unity. We stopped our simulation after we halved $\Delta F$ 
$20$ times. Therefore, the final $\Delta F$ is $2^{-20}$. 
The magnetization space $(m_{\perp},m_{\parallel})$ was divided into 
$10,000$ bins with the area of $0.01\times0.01$, and free-energy 
was calculated for each bin which satisfies the inequalities 
$0\le (m_{\perp}^2+m_{\parallel}^2)^{1/2}\le 0.85$. The histogram of two magnetizations 
$H(m_{\perp},m_{\parallel})$ was checked every $10,000$ MC steps. We regarded the histogram as flat when 
$H(m_{\perp},m_{\parallel})$'s for all the magnetizations are not less than 
$80\%$ of the average value of the histogram. 
The SCO method was applied only to dipolar interactions. 
A potential switching procedure in the SCO method was performed every $10$ MC steps. 
We performed the free-energy measurement for $10$ different runs with different initial 
states and random number sequences to estimate the means and errorbars of the free-energy. 
The estimated error bars were rather small. They were smaller than $0.03~J$ 
for all of the data we will show hereafter. 

\begin{figure}[t]
\begin{center}
\includegraphics[width=\columnwidth]{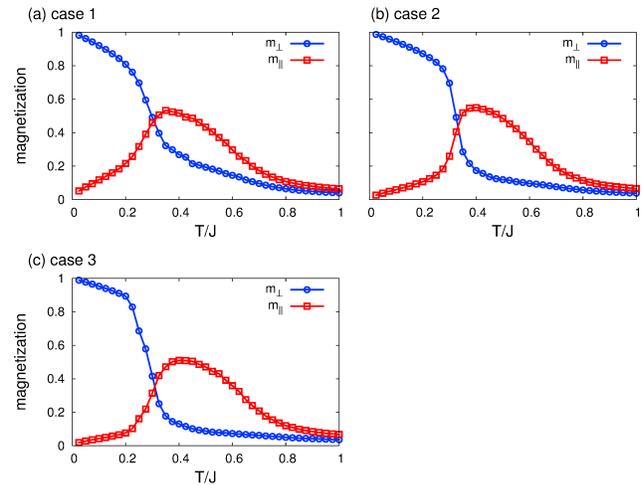}
\end{center}
 \caption{(Color online) The temperature dependences of the perpendicular magnetization $m_{\perp}$ 
(open circles) and in-plane magnetization $m_{\parallel}$ (open squares) in the case 1 (upper left), 
2 (upper right), and 3 (lower left) in Table~\ref{param}. In the measurement, the temperature was 
gradually cooled from $2.0~J$ to $0.025~J$ in steps of $\Delta T=0.025~J$. The system was kept for 
$100,000$ MC steps at each temperature. The data of the latter $50,000$ MC steps were used to 
calculate the thermal averages of the two magnetizations. The average was also 
taken over 10 different runs with different initial states and random number sequences.
}
 \label{mag-t}
\end{figure}

\begin{figure}[t]
\begin{center}
\includegraphics[width=\columnwidth]{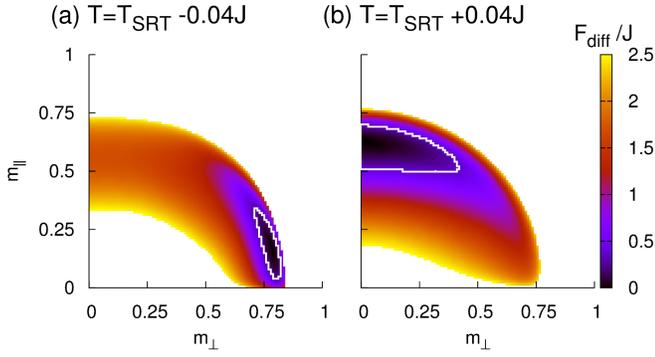}
\end{center}
 \caption{(Color online) Free-energy difference $F_{\rm diff}$ in the case 2 is plotted 
as a function of $(m_{\perp},m_{\parallel})$ for two different temperatures. 
$F_{\rm diff}$ is defined by eq.~(\ref{eqn:Def_DeltaF}). 
The left and right panels show 
the data for $T=T_{\rm SRT}-0.04~J$ and those for $T=T_{\rm SRT}+0.04~J$, 
respectively. $T_{\rm SRT}$ is $0.33~J$. 
In both figures, $F_{\rm diff}$ is plotted only for a region 
in which $F_{\rm diff}$ is smaller than $2.5~J$. The low free-energy region 
defined by the inequality $F_{\rm diff} \le T$ is denoted by a solid white line. 
The average was taken over $10$ different runs. 
} \label{free1}
\end{figure}

\section{Results}
\label{sec:results}
To check whether this model exhibits the SRT or not, we first measured 
the perpendicular and in-plane magnetizations as a function of the temperature
in the three cases 1, 2, and 3 in Table~\ref{param}. Figure~\ref{mag-t} shows the result. 
We hereafter set the Boltzmann constant $k_{\rm B}$ to unity and use $J$ as a unit of temperature. 
In this measurement, the temperature was gradually cooled from $2.0~J$ 
to $0.025~J$ in steps of $\Delta T=0.025~J$, and the thermal averages 
of the two magnetizations were measured at each temperature. 
As the temperature decreases, $m_{\parallel}$ starts to increase rapidly 
around $T\approx 0.8~J$. On the contrary, such rapid increase does not occur 
in $m_{\perp}$ around this temperature. Therefore, $m_{\parallel}$ is dominant 
at intermediate temperatures. However, $m_{\parallel}$ starts to drop 
around $0.4~J$, and $m_{\perp}$ overtakes $m_{\parallel}$ around $0.3~J$. 
These data clearly show that the present model exhibits the SRT. 
We also see that the SRT temperature $T_{\rm SRT}$, 
which is defined by the temperature where $m_{\perp}$ is equal to $m_{\parallel}$, 
is around $0.3~J$ in all of the three cases. This means that 
$T_{\rm SRT}$ mainly depends on the ratio $C_{\rm d}/C_{\rm u}$. 
Recall that the ratio is fixed to $0.14$ in all the four cases (see Table~\ref{param}). 
We also performed MC simulation in the case~4, 
and found that a two-domain structure emerges in this case. 
Because the purpose of the present study is to investigate the SRT in single-domain 
magnetic films, we hereafter show the results of the three cases 1-3. 
Since the size of the case 3 is close to that of the case 4, 
the three cases 1-3 cover almost the whole range of the size 
in which a single-domain structure is retained.

One may consider that a pseudogap does not exist in this model 
because neither $m_{\perp}$ nor $m_{\parallel}$ is zero at $T_{\rm SRT}$. 
However, this result is not incompatible with the presence of 
a pseudogap due to the dynamical origin. 
It should be recalled that {\it each component of magnetization} $m_{\rm \mu}$ ($\mu=x,y,z$), 
which can be both positive and negative, disappears at $T_{\rm SRT}$. 
When spins are aligned in the same direction and the direction changes with time, 
the time average of $m_{\rm \mu}$ becomes zero. However, the time averages of 
$m_{\perp}$ and $m_{\parallel}$ do not become zero because they are always positive 
by definition.

\begin{figure}[t]
\begin{center}
\includegraphics[width=\columnwidth]{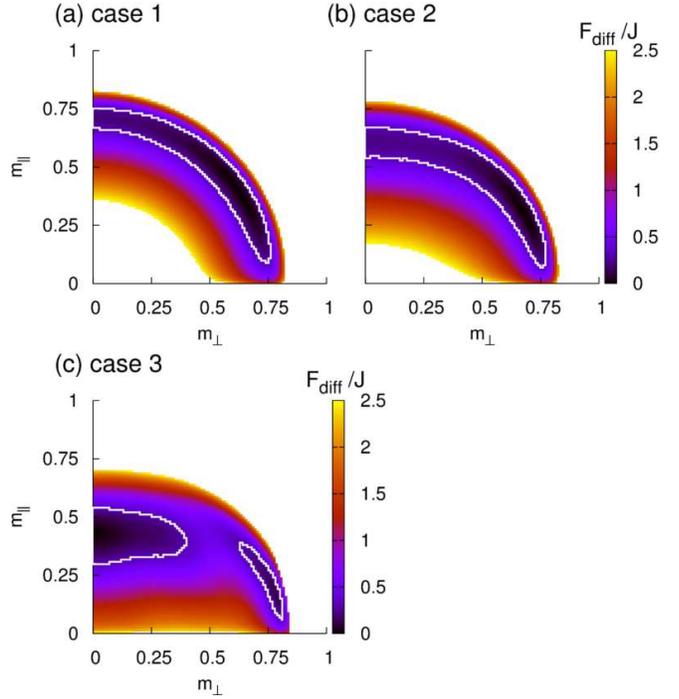}
\end{center}
 \caption{(Color online) Free-energy difference $F_{\rm diff}$ at $T_{\rm SRT}$ is plotted 
as a function of $(m_{\perp},m_{\parallel})$ for the case 1 (upper left), 2 (upper right), 
and 3 (lower left). $F_{\rm diff}$ is plotted only for a region 
in which $F_{\rm diff}$ is smaller than $2.5~J$. The low free-energy region 
defined by the inequality $F_{\rm diff} \le T$ is denoted by a solid white line. 
The average was taken over $10$ different runs. 
} \label{free2}
\end{figure}

We next measured the free-energy $F(\beta;m_{\perp},m_{\parallel})$ defined by eq.~(\ref{def_f}). 
In Fig.~\ref{free1}, free-energy difference $F_{\rm diff}$ in the case 2 
is plotted as a function of $(m_{\perp},m_{\parallel})$, 
where $F_{\rm diff}$ is defined by 
\begin{equation}
F_{\rm diff}(\beta;m_{\perp},m_{\parallel}) \equiv F(\beta;m_{\perp},m_{\parallel})-F_{\rm min}(\beta),
\label{eqn:Def_DeltaF}
\end{equation}
and 
\begin{equation}
F_{\rm min}(\beta)\equiv \min\nolimits_{(m_{\perp},m_{\parallel})} F(\beta;m_{\perp},m_{\parallel}).
\end{equation}
Figures~\ref{free1}(a) and \ref{free1}(b) show 
the data for $T<T_{\rm SRT}$ and those for $T>T_{\rm SRT}$, respectively. 
We see that $F_{\rm diff}$ for $T<T_{\rm SRT}$ has a global minimum 
around $(m_{\perp},m_{\parallel})\approx (0.75,0.15)$. In contrast, when $T>T_{\rm SRT}$,  
$F_{\rm diff}$ has a global minimum around $(m_{\perp},m_{\parallel})\approx (0,0.6)$. 
The location of the global minimum shifts discontinuously as the temperature is 
changed across $T_{\rm SRT}$. In this sense, we can regard the SRT as a first-order 
transition. This discontinuous shift of the global minimum explains the reason why 
$m_{\perp}$ and $m_{\parallel}$ in Fig.~\ref{mag-t} change abruptly around $T_{\rm SRT}$. 
This result is consistent with the previous results 
that single monolayer magnetic films with a single-domain structure exhibit 
a first-order transition~\cite{Hucht95,SantamariaDiep00,MoschelUsadel95}. 
A similar discontinuous shift in the global minimum 
was observed in the other two cases $1$ and $3$. 

\begin{figure}[t]
 \begin{center}
\includegraphics[width=\columnwidth]{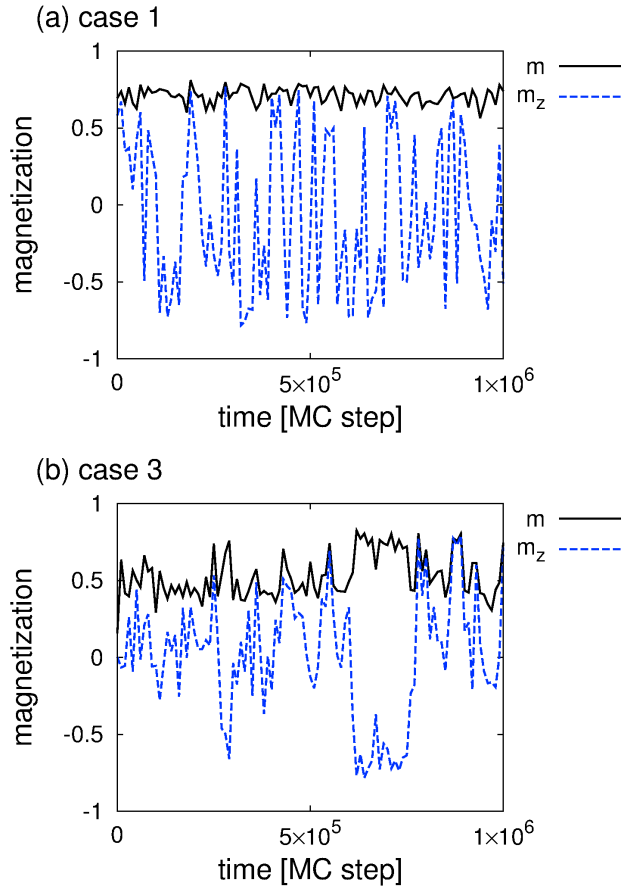}
 \end{center}
 \caption{(Color online) The amplitude of magnetization $m\equiv (m_{\perp}^2+m_{\parallel}^2)^{1/2}$ 
and the $z$-component of magnetization $m_z$ measured at $T_{\rm SRT}$ 
are plotted as a function of time for the two cases 1 (upper panel) and 3 (lower panel).}
 \label{time-course}
\end{figure}

We next focus our attention to the free-energy structure at $T_{\rm SRT}$. 
In Fig.~\ref{free2}, we show $F_{\rm diff}$ for all of the three cases. 
There are several differences in the free-energy structure among 
the three cases. When $C_{\rm d}$ and $C_{\rm u}$ are small (case 1), 
the free-energy has a circular structure. In the case 2, this free-energy 
structure is slightly deformed into a elliptical one. 
For large $C_{\rm d}$ and $C_{\rm u}$ (case 3), the deformation proceeds further, 
and the free-energy has two local minima and a free-energy barrier in between. 
However, a common feature among the three free-energy structures is that 
a low free-energy region defined by the inequality $F_{\rm diff} \le T$, 
which is denoted by a solid white line in Fig.~\ref{free2}, 
widely spreads in the $(m_{\perp},m_{\parallel})$ space. 
In the cases 1 and 2, low free-energy region involves both 
the perpendicularly magnetized state with $m_{\parallel}\approx 0$ 
and the in-plane magnetized state with $m_{\perp}\approx 0$. 
There is no (or vanishingly small) barrier between the two states. 
In the case 3, there is a free-energy barrier like an ordinary first-order 
phase transition. However, the barrier is rather small. 
Because single-domain magnets are known to exhibit large fluctuations in magnetization 
if the energy barrier is small in comparison with $k_{\rm B}T$~\cite{SPbehavior}, 
we can expect that magnetization at $T_{\rm SRT}$ heavily fluctuates with time. 

To verify weather this reasoning is correct or not, we performed a standard MC simulation 
at $T_{\rm SRT}$ and measured the amplitude of magnetization $m\equiv (m_{\perp}^2+m_{\parallel}^2)^{1/2}$ 
and the $z$-component of magnetization $m_z$. Figure~\ref{time-course} shows the result. 
The data in the case 1 and those in the case 3 are shown in Figs.~ \ref{time-course}(a) 
and \ref{time-course}(b), respectively. We see that $m_z$'s in both the cases 
heavily fluctuate with time, as expected. A similar behavior was observed 
in the other two components of magnetization $m_x$ and $m_y$. 
However, the amplitude of magnetization $m$ is always non-zero because 
$m$ is non-zero in the low free-energy region of $F_{\rm diff}$. 
Especially, in the case 1, $m$ in the low free-energy region is almost constant 
because $F_{\rm diff}$ has a circular structure. This is the reason why fluctuations 
in $m$ are rather small in the case 1. These results show that, at the SRT temperature, 
spins are aligned in the same direction, and the direction changes with time. 
In other words, the system is in a superparamagnetic state at $T_{\rm SRT}$. 
Therefore, the pseudogap in magnetization emerges around the SRT temperature 
because of dynamical fluctuations in magnetization which are inherent 
in the superparamagnetic state.

\begin{figure}[t]
 \begin{center}
\includegraphics[width=\columnwidth]{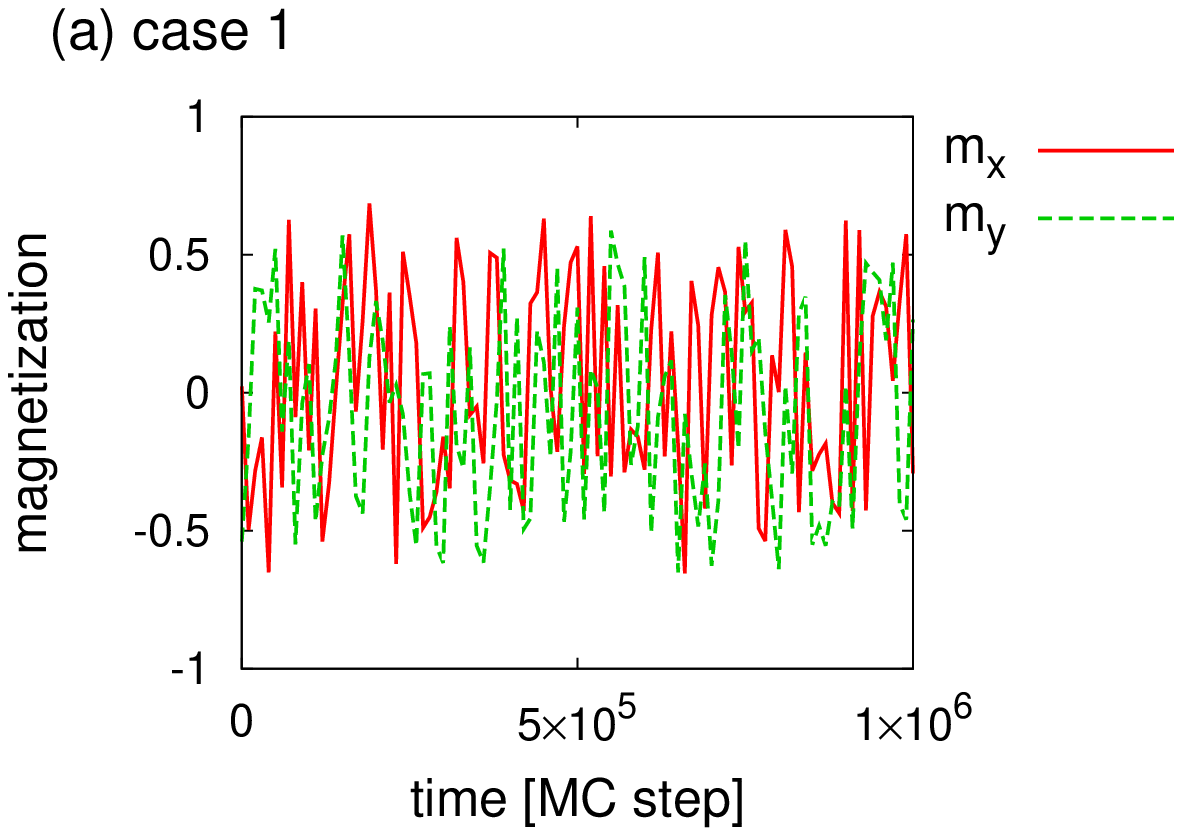}
\includegraphics[width=\columnwidth]{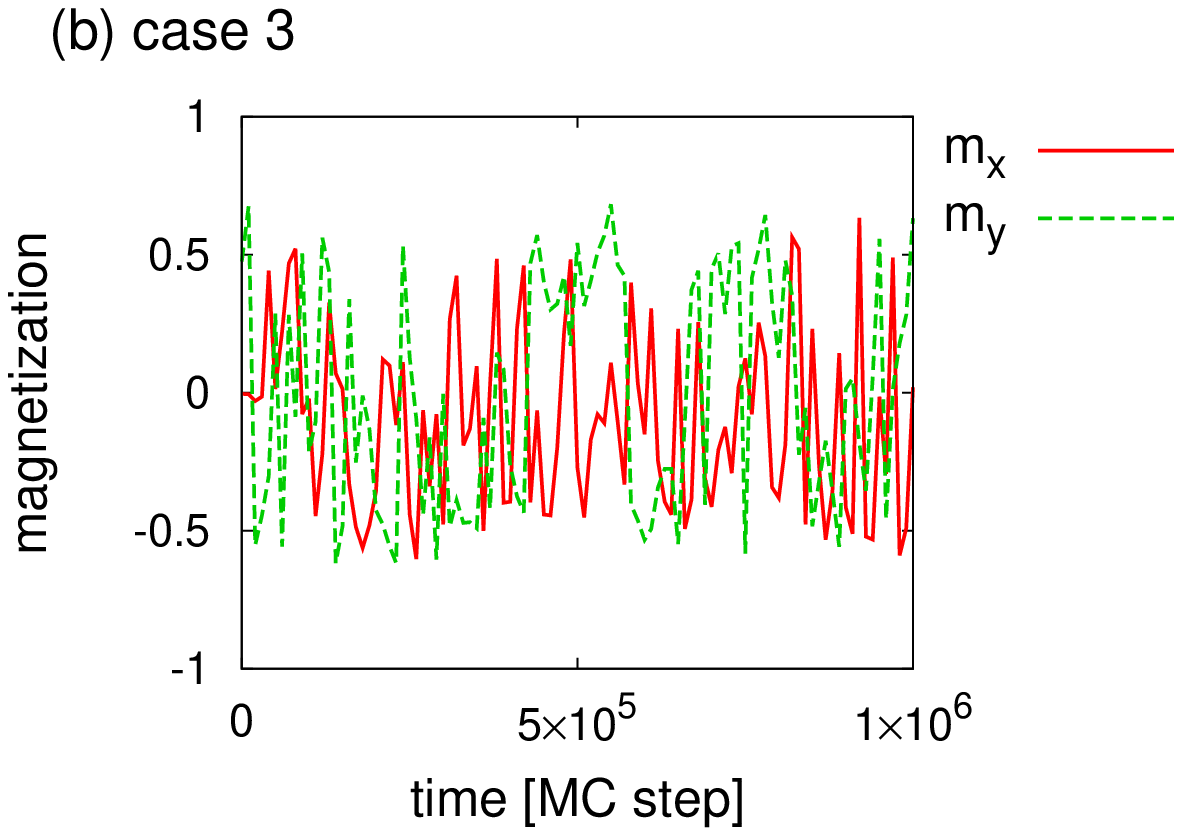}
 \end{center}
 \caption{
(Color online) In-plane components of magnetizations $m_x$ and $m_y$ measured at $T=0.4~J$ are plotted as a function 
of time for the two cases 1 (upper panel) and 3 (lower panel).}
\label{time-course2}
\end{figure}

One may consider that, even if the system is in a in-plane magnetized state and 
$m_\parallel\ne 0$, the time averages of two magnetizations $m_x$ and $m_y$ become zero 
if magnetization freely fluctuates in the plane of the film. To examine whether 
such fluctuations occur or not, we measured the time evolutions 
of $m_x$ and $m_y$ at $T=0.4$~J for the two cases 1 and 3. 
Note that the system is in a in-plane magnetized state at this temperature 
(see Fig.~\ref{mag-t}). The result is shown in Fig.~\ref{time-course2}. 
We see that $m_x$ and $m_y$ heavily fluctuate with time. This result means that 
the four-fold magnetic anisotropy on square lattice, which is induced by magnetic dipolar interactions, 
is not sufficient to prevent in-plane fluctuations of magnetization. 
To prevent these fluctuations,  
some additional anisotropies such as the shape magnetic anisotropy 
and induced magnetic anisotropy which stabilize magnetization to a direction in the plane
are necessary.


Lastly, it may be worth pointing out the similarities and differences between our model and the normal 
ferromagnetic model which involves only exchange interactions, i.e., our model 
with $C_{\rm d}=C_{\rm u}=0$. A wide spread of low free-energy region 
in the magnetization space and large fluctuations 
in magnetization are also observed in the normal ferromagnetic model. However, 
because exchange interactions are isotropic, such behavior is observed at {\it any} temperatures 
in this model. On the contrary, this behavior is observed only around $T_{\rm SRT}$ in our model. 
This is the difference between the two models. Because the pseudogap emerges 
only around $T_{\rm SRT}$ in experiments, the normal ferromagnetic model is not proper 
to describe the SRT. 

\section{Conclusions}
\label{sec:conclusions}
We investigated the spin reorientation transition (SRT) 
in ultrathin magnetic films by Monte-Carlo simulations. We concentrated ourselves 
on the case that the lateral size of the film is relatively small and the system has a 
single-domain structure. As a result of free-energy measurement, 
we found that the system is in a superparamagnetic state at $T_{\rm SRT}$
and the disappearance (pseudogap) of magnetization around $T_{\rm SRT}$ emerges 
due to dynamical fluctuations in magnetization. This observation is in contrast to that 
in large ultrathin magnetic films that the pseudogap is caused by a static magnetic structure 
with many complex domains.

\section*{Acknowledgment}
The authors would like to thank Professor K. Sasaki for valuable discussions and comments.


\begin{thebibliography}{99}
\bibitem{note1}
In ref.~\citen{Carubelli08}, the authors used the same Hamiltonian as eq.~(\ref{Hami}) 
and set the ratio $C_{\rm d}/J$ to be $0.33$. By assuming that the material parameters 
$A$ and $J_{\rm s}$ are the same as those used in the calculation shown in Table~\ref{param},
the mesh size $\delta$ is calculated to be about $10$~nm from eq.~(\ref{eqn:parameters}) and the ratio $C_{\rm d}/J$. 
Because the lattice size is $40\times 40 \times 1$, the corresponding system size 
is estimated to be $400~\textrm{nm}\times 400~\textrm{nm}\times 10~\textrm{nm}$.
\bibitem{SPbehavior}
For example, the relaxation time $\tau$ of a single-domain nano-particle 
with uniaxial anisotropy is well described by the Arrhenius law
\[
\tau=f_0^{-1}\exp[\Delta B/k_{\rm B}T],
\]
where $\Delta B$ is the energy barrier between the two stable states and 
$f_0$ is the attempt frequency which is of the order of $10^9\sim 10^{11}~{\rm s}^{-1}$.
\end{thebibliography}
\end{document}